\documentclass[rsi %
 aip,
 amsmath,amssymb,
 reprint,%
]{revtex4-1}

\usepackage{graphicx}
\usepackage{dcolumn}
\usepackage{bm}

\usepackage[utf8]{inputenc}
\usepackage[T1]{fontenc}
\usepackage{mathptmx}
\usepackage{etoolbox}
\usepackage{siunitx}
\UseRawInputEncoding

\makeatletter
\def\@email#1#2{%
 \endgroup
 \patchcmd{\titleblock@produce}
  {\frontmatter@RRAPformat}
  {\frontmatter@RRAPformat{\produce@RRAP{*#1\href{mailto:#2}{#2}}}\frontmatter@RRAPformat}
  {}{}
}%
\makeatother
\begin{document}

\preprint{AIP/123-QED}

\title[Electrospray Propulsion Time-of-Flight Secondary Ion Mass Spectrometry Diagnostic]{Electrospray Propulsion Time-of-Flight Secondary Ion Mass Spectrometry Diagnostic}
\author{Giuliana C. Hofheins}
\email{gch72@cornell.edu}
\author{Zach Ulibarri}%
\author{Elaine M. Petro}
\affiliation{Cornell University, Ithaca, NY, 14850, USA}
\homepage{https://www.astralab.mae.cornell.edu}

\date{\today}

\begin{abstract}
The design and capability of a novel time-of-flight secondary ion mass spectrometry electrospray propulsion diagnostic is presented to investigate secondary species emission from surface impingement of high-velocity, energetic molecular ion plumes. Designed on the basis of traditional Secondary Ion Mass Spectrometry (SIMS) principles, this diagnostic provides information on the relative intensity and chemical composition of secondary species given electrospray operational parameters like incident angle, primary ion energy, and target surface composition. The system consists of an externally-wetted tungsten ion source operating with room temperature ionic liquid propellant, a target with a secondary species extraction mesh, and a time-of-flight mass spectrometer featuring an electrostatic deflection gate and a multichannel plate detector. Results show that energetic primary plume impacts with metallic surfaces induce molecular secondary ion emission in both positive and negative polarities. Likely sources of these secondary ions are considered – including hydrocarbon contamination of the target surface and charged fragments of the molecular primary ionic liquid ions. For electrospray propulsion, these secondary species contribute not only to lifetime limiting processes intrinsic to thruster operation like impingement and thus degradation of electrodes and emitters, but also contribute to facility effects corrupting ground-based testing and thruster flight qualification. 
\end{abstract}

\maketitle

\section{\label{sec:Introduction}Introduction}
Electric space propulsion systems produce thrust via accelerating propellant through a combination of electric and magnetic fields. The appeal of electric propulsion (EP) lies in that propellant exhaust velocity is determined by input power rather than intrinsic chemical bond energy, thus enabling highly fuel-efficient systems for lower spacecraft propellant mass fractions \cite{goebel2023fundamentals}. Electrospray micropropulsion thrusters are a subset of electric propulsion systems that utilize electrostatic fields to extract and accelerate charged particles from a liquid propellant on a either a sharp emitter tip or capillary meniscus \cite{lozano_electrospraypropulsion_2010}. When a potential difference on the order of a few kilovolts is applied between the emitter a downstream extractor electrode, a Taylor cone is formed from a combination of surface tension, electric, and hydrodynamic stresses \cite{petro_characterization_2020, gallud2022emission}. Ion emission is thus induced through field-emission evaporation, where ions are subsequently accelerated to velocities exceeding 10,000 m/s \cite{petro_characterization_2020}. This process forms polydisperse plumes ($\sum_{i=1} q_i/m_i$) that generate thrust on the level of nano to a few millinewtons per emitter \cite{petro_multiscale_2022}. Plume composition ranges from pure ion emission \cite{lozano_ionic_2005} to larger multiply-charged droplets \cite{st7_hruby} depending on emitter type and operating modes. 

\begin{figure}
  \centering  \includegraphics[width=1\linewidth]{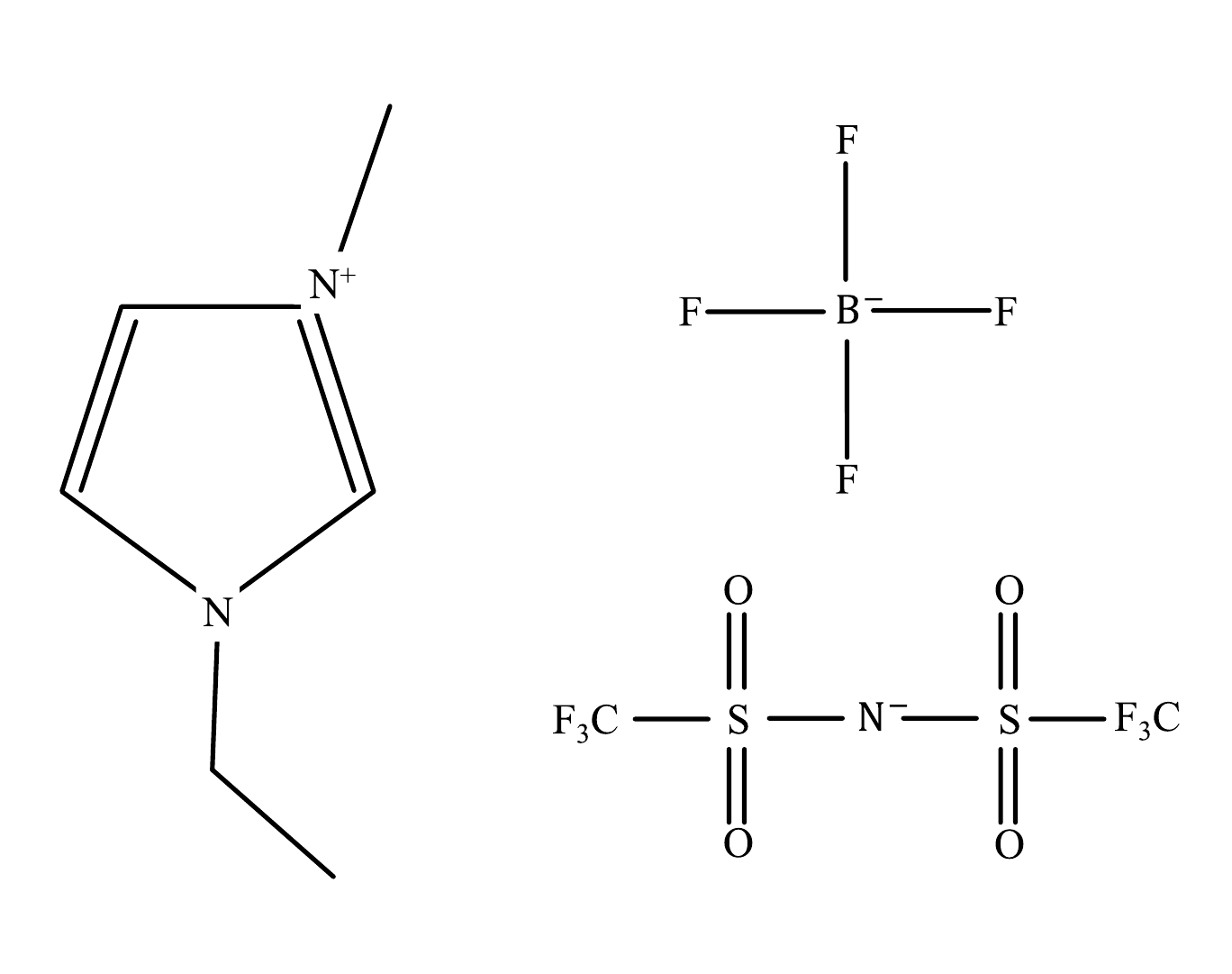}
  \caption{Structural formulas for common ionic liquid propellants, the cation 1-ethyl-3-methylimidazolium (EMI$^+$), and anions tetraflouroborate (BF$_4^-$), and bis(trifluoromethylsulfonyl)imide (Im$-$).}
  \label{fig: ionic liquids}
\end{figure}

A subset of electrospray devices utilize chemically complex molten salt propellants known as room temperature ionic liquids (RTILS), comprised of organic cations and anions. For example, common propellants include 1-ethyl-3-methylimidazolium tetrafluoroborate (EMI-BF$_4$) and 1-ethyl-3-methylimidazolium bis(trifluoromethylsulfonyl)imide (EMI-Im) as shown in Fig.\@ \ref{fig: ionic liquids}, where the EMI$^+$ cation is composed of an imidazolium ring with alkyl functional groups. These propellants are attractive for electrospray propulsion devices given that they are nonvolatile and can be operated in either polarity,  eliminating the need for an external neutralizer \cite{lozano_nanoengineered_2015}. The combination of these properties make RTIL electrospray thrusters attractive for missions requiring efficient, compact, and low-mass and power propulsion systems. Such examples include small satellite propulsion \cite{mier-hicks_electrospray_2017}, as well precise spacecraft control like the LISA Pathfinder technology demonstration mission which tested EMI-Im colloid thrusters \cite{ziemerST7}.

\begin{table}[]
    \centering
    \begin{tabular}{cccc}
        \hline
        \hline
        Polarity & Plume Species  & Order & Molecular Mass (amu) \\
        \hline
       $+$ & EMI$^+$(EMI-BF$_4$)$_n$ &  n = 0 & 111 \\
        & & n = 1  & 309 \\
        & & n = 2  & 507 \\
        & & & \\
        $-$ & BF$_4^-$(EMI-BF$_4$)$_n$ &  n = 0 & 87 \\
        & & n = 1 & 285 \\
        & & n = 2 & 483 \\
         \hline
         \hline
    \end{tabular}
    \caption{Primary EMI-BF$_4$ ionic liquid ion plume species dependent on firing polarity, including primary monomers (n = 0) and cationated and anionated neutrals (n > 0). }
    \label{tab:my_label}
\end{table}

However, complex organic RTILs have vastly different chemical properties compared to inert noble gas electric propulsion propellants like xenon and krypton employed in ion and Hall thrusters \cite{goebel2023fundamentals}. While noble gas ion-surface interactions have decades of research due their nearly ubiquitous use among EP systems, the surface interactions of complex ions are hypothesized to be much different \cite{kolasinski_fundamental_2007}. For example, they may deposit at low energies or form new fragment product ions at high energies \cite{bendimerad_molecular_2022}. Both experiments \cite{krejci_emission_2017} and simulation \cite{petro_multiscale_2022} indicate that particles from the plume impact thruster and spacecraft surfaces like the extractor electrode, resulting in propellant accumulation and eventual device failure \cite{thuppul_lifetime_2020}. However, there is a wide gap in the knowledge of the fundamental physics of these molecular ion-surface collisions at the nanoscale and how they ultimately impact thruster lifetime and performance. Moreover, these systems are tested and validated in ground-based vacuum testing environments where secondary species from plume-surface impacts are known to affect diagnostics \cite{ma_plume_2022, uchizono_role_2021}. Thus, it is important to know the chemical composition of secondary species not only for system modeling and lifetime considerations -- but for facility effects of electrospray testing. The uncertainty of collision products makes it difficult to predict, correct for, and model these discrepancies.

Both numerical simulations and experimental investigations have been conducted to characterize electrospray plume-surface interactions. Molecular dynamics (MD) studies have been performed both with non-reactive \cite{takahashi_tungsten_2009, cidoncha_modeling_2022} and reactive force fields \cite{bendimerad_molecular_2022} with EMI-BF$_4$ impacting various target surfaces. Bendimerad, et al.\@ \cite{bendimerad_molecular_2022} produced pseudo-mass spectra of EMI-BF$_4$ impacting a potential wall with a reactive force field showing the relation between relative intensity of fragmentation products and impact energy. 
Experimental efforts from Uchizono \textit{et al.}\@ \cite{uchizono2022positive, uchizono_dissertation_nodate} and Klosterman \textit{et al.} \cite{klosterman_ion-induced_2022} have quantified secondary species yields and \citet{uchizono_role_2021} has examined the role of secondary species on experimental testing. Shaik \textit{et al.}\@ \cite{shaik_characterization_2024} employed a residual gas analyzer (RGA) to identify secondary species from an electrospray thruster plume impinging on stainless steel. However, both MD and experimental efforts lack in determining the chemical composition of charged secondary species that arise from plume-surface interactions across the permutation of thruster operation variables (target surface composition, primary beam energy and current, background pressure).

\begin{figure}[b]
  \centering
  \includegraphics[width=1\linewidth]{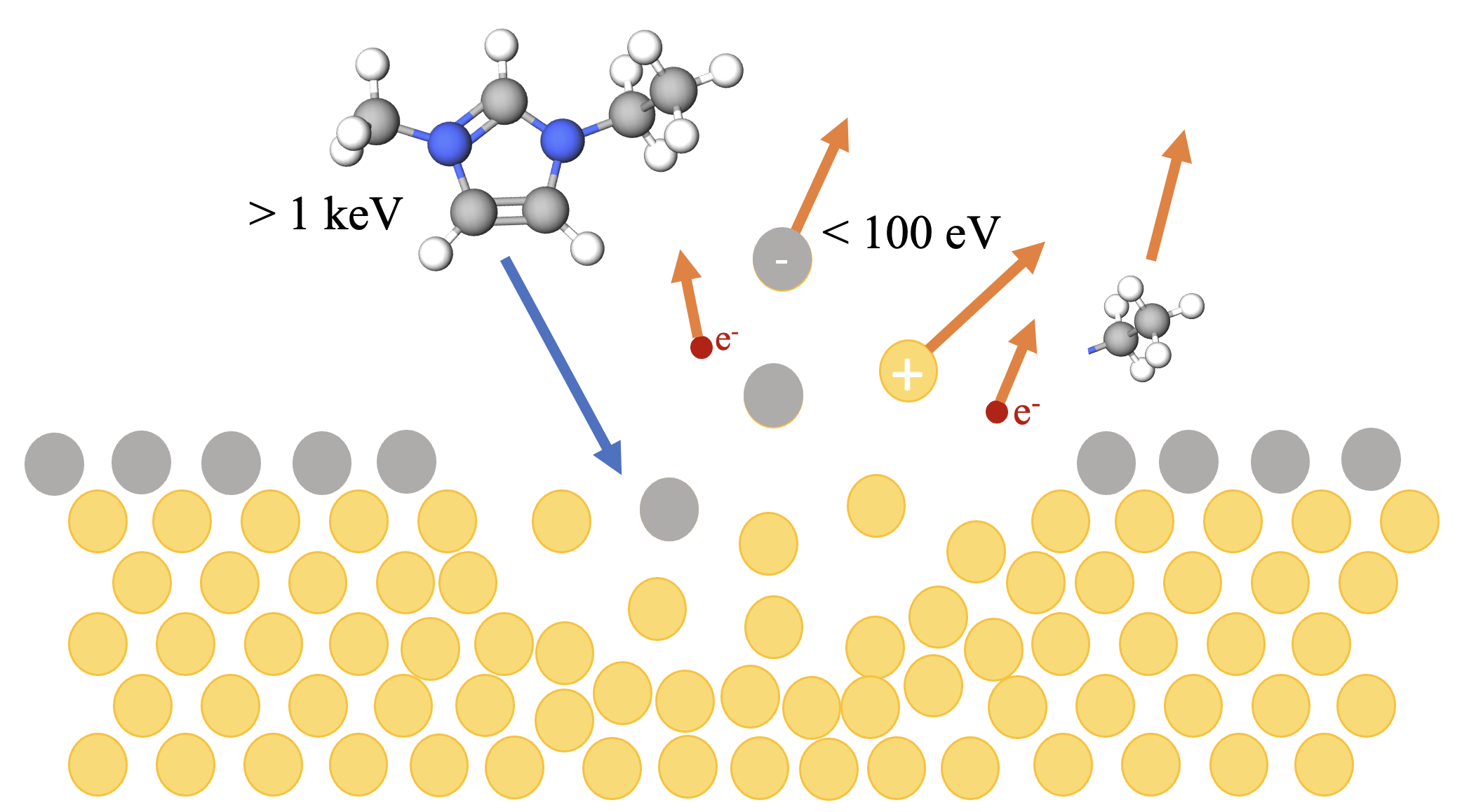}
  \caption{Simplified kinetically-assisted potential sputtering of a metallic target with a representative adsorbed monolayer (oxides, hydrocarbons, etc).}
  \label{fig: kinetic_sputtering}
\end{figure}

Time-of-flight secondary ion mass spectrometry (TOF-SIMS) is a standard surface composition characterization technique, commonly used either in static mode to non-destructively characterize the outermost monolayer of a sample, or in dynamic mode to probe depth-based composition \cite{van_der_heide_secondary_2014}. The general premise is to fire a high energy ion beam (1-100 keV) at an incidence angle upon the surface of interest. This induces sputtering, where a fraction of sputtered species exist as low energy ions \cite{van_der_heide_secondary_2014}. To form the secondary ion beam, these sputtered secondary ions are extracted and accelerated down a time-of-flight tube to a detector, where a mass spectrum of secondary ions is generated based on conservation of energy principles \cite{lozano_ionic_2005}. The resulting spectrum gives insight into the atomic and molecular features of the sample of interest. Canonically, SIMS is employed with high energy large atomic primary ion beams [Cs+, Ga+, etc], and sputtering is most generally modeled via purely kinetic momentum transfer from isotropic collision cascades \cite{van_der_heide_secondary_2014}. The energy of the sputtered population is generally modeled as Maxwell-Boltzmann distributions with peak energies below 10 eV and tails extending to a few hundred eV \cite{van_der_heide_secondary_2014}. Alternatively, current research shows that small molecular primary ion beams (Bi$_n^+$, SF$_5^+$, In$_n^+$, etc) induce kinetically-assisted sputtering as shown in Fig.\@ \ref{fig: kinetic_sputtering} \cite{van_ham_comparison_2005, fuoco_surface_2001, townes_mechanism_1999}.  This type of sputtering removes surface atoms, molecules, or ions from overlapping collisions from a singular initial collision event that induces thermal evaporation of substrate atoms and ions \cite{van_der_heide_secondary_2014}. This method has not only proven increased sputter yields but also the ability to detect intact large organic molecules from the target surface \cite{gillen_preliminary_1998}.

The difference between a traditional TOF-SIMS primary ion beam and an electrospray beam lies in both the complex organic molecular composition of RTIL electrospray plumes and the typical experimental vacuum environments. These complex plumes give rise to multiple processes like deposition and fragmentation both in the plume and upon impact \cite{petro_multiscale_2022, bendimerad_molecular_2022}. However, work by Fujiwara, et al \cite{fujiwara_time--flight_2014} has proven that ionic liquids (EMI-Im) can be effective as TOF-SIMS primary ion beams. In addition, commercial TOF-SIMS are ultra-high vacuum systems ($<$ 10$^{-9}$ Torr) in order to mitigate background gas adsorbing onto the target surface and interfering with the signal \cite{van_der_heide_secondary_2014}. Electrospray systems are tested in a range of vacuum environments, with some facility pressures orders of magnitude higher than that of commercial TOF-SIMS. This implies that in nominal test environments, an electrospray plume is impacting not only bare surfaces but also those coated with a monolayer of adsorbed background gas contaminants and oxide/hydrocarbon layers \cite{klosterman_ion-induced_2022}, thus further convoluting device performance and diagnostic evaluation.

Electrospray plumes impact surfaces at high energies ( $>$ 1 keV) and thus induce sputtering, with some portion of the sputtered population composed of secondary ions. These secondary ions will interact with the electrostatic fields in thruster operation and testing, and thus it is important to characterize their chemical composition. To accomplish this, a novel ESI TOF-SIMS diagnostic has been developed to characterize plume-surface interactions of electrospray systems. This diagnostic will inform the relevant impact secondary ion products from given electrospray operating conditions -- including impact energy, propellant choice and target surface composition. The design of the experimental diagnostic is detailed in Section \ref{sec: ESI TOF-SIMS Design}, with secondary ion mass spectra of both polarities from an ionic liquid primary beam impacting metallic samples presented and discussed in Section \ref{sec: Results}. Finally, conclusions and future work is detailed in Section \ref{sec: conclusion}.

\section{ESI TOF-SIMS Design}
\label{sec: ESI TOF-SIMS Design}

\begin{figure*}
\includegraphics[width=1\linewidth]{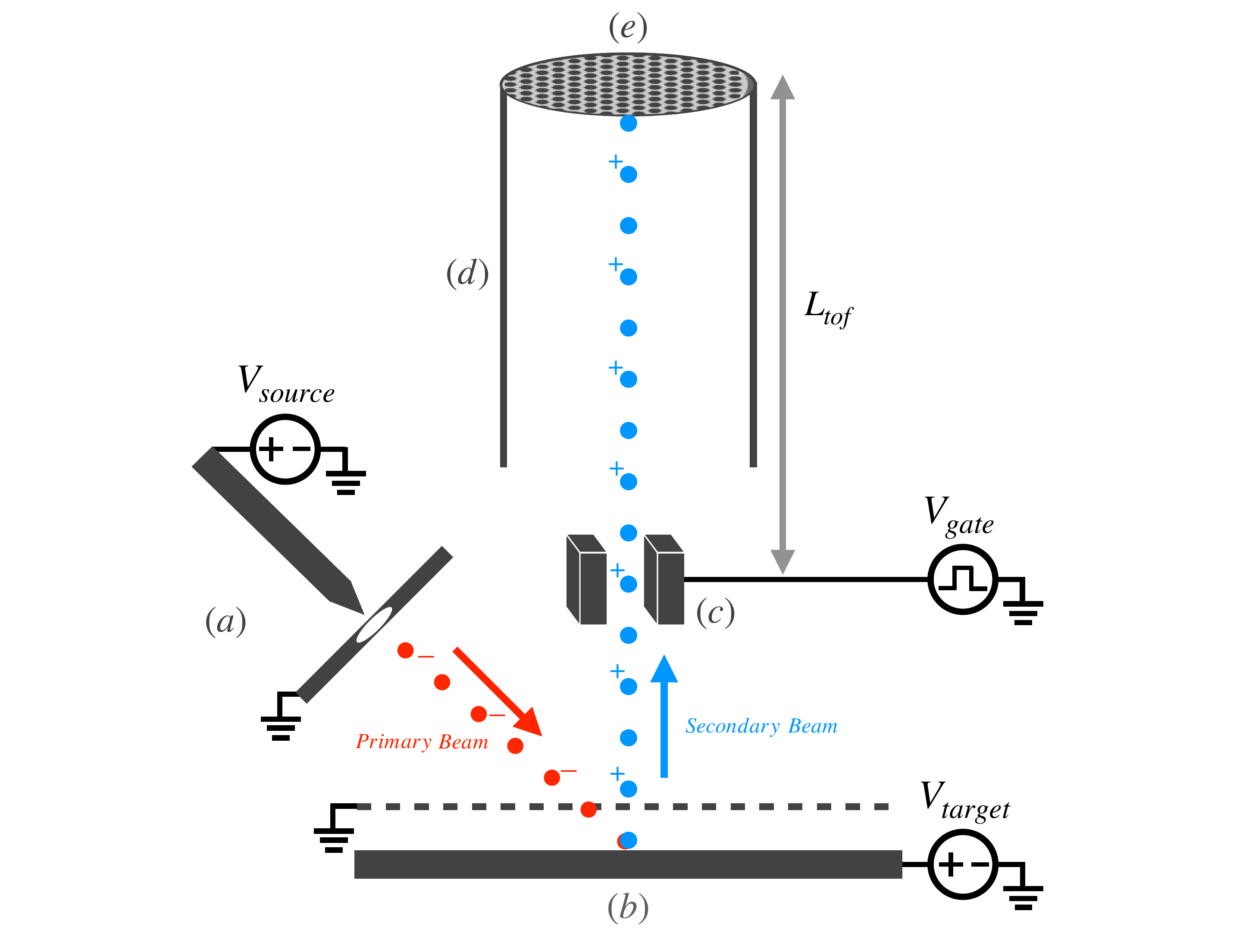}
\caption{\label{fig: setup} The source (a) directs the primary beam at the target (b), where secondary ion emission is induced and directed toward the electrostatic gate (c) and time-of-flight system (d) with an MCP detector (e). To collect secondary ions, the polarities of the ion beams were reversed. }
\end{figure*}
 
 The experimental design of the electrospray TOF-SIMS diagnostic is shown in Fig. \ref{fig: setup}.
 The design consists of (a) an externally wetted tungsten single emitter primary electrospray ion source operating with EMI-BF$_4$ propellant, (b) a high-voltage target assembly with a secondary ion acceleration grid, (c) an electrostatic deflection gate, (d) a time-of-flight tube and (e) a microchannel plate detector. This diagnostic was operated in a 2 ft diamter cylindrical vacuum chamber with rotary vane roughing pump and turbomolecular vacuum pump to achieve pressures between \num{5e-5} Torr to \num{1e-6} Torr. A primary molecular ion plume is first produced via the tungsten electrospray ion source. This plume impacts a target surface at a given incidence angle with an impact energy dependent on the source ($V_{source}$) and target ($V_{target}$) potentials. The high energy impacts induce removal of secondary electrons, atoms, molecules and ions from the target surface \cite{uchizono_dissertation_nodate}. In addition, processes like collision-induced dissociation likely contribute to secondary population. \cite{uchizono_role_2021,bendimerad_molecular_2022}. These low energy ($<$ 100 eV) secondary ions, electrons, and neutrals are ejected from the target surface typically modeled as a cosine angular distribution from the surface normal \cite{van_der_heide_secondary_2014,uchizono_dissertation_nodate}. Secondary ions are selectively extracted from the sputter cloud from a potential difference between the high voltage target surface and a grounded secondary ion acceleration grid. This single polarity, approximately monoenergetic secondary ion beam passes through a pulsed high voltage electrostatic deflection gate and down a linear time-of-flight tube. The secondary ions impact a microchannel plate detector, and a time-of-flight curve and mass spectrum is produced based on the flight time of a given species.  The individual components of the electrospray TOF-SIMS system are described further in the following sections.

\subsection{Ion Source}
\begin{figure}[h!]
\includegraphics[width=1\linewidth]{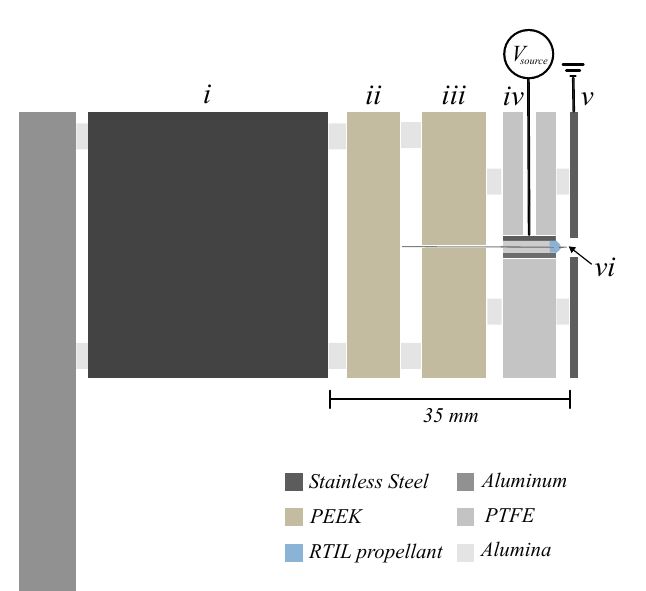}
\caption{\label{fig: source} A side profile cross section of the ion source, with (i) goniometer base, PEEK baseplates (ii, iii), PTFE reservoir plate (iv) with stainless steel reservoir, stainless extractor (v) and externally-wetted tungsten needle emitter (vi).}
\end{figure}

A cross-section of the ion source employed in the diagnostic is shown in Fig. \ref{fig: source}. The ionic liquid electrospray ion source consists of a 0.5 mm diameter tungsten needle with a tip electrochemically etched as described in Lozano, \textit{et al}.\cite{lozano_ionic_2005} to approximately a 2.9 $\mu m$ radius of curvature as determined by approximate circular fits with a scanning electron microscope. This tungsten emitter is affixed to a dual stage goniometer (OptoSigma GOH-40B35) via polyetheretherketone (PEEK) plates and alumina spacers (Kimball physics) in order to place the needle tip at the goniometer rotation center while maintaining electrical isolation of the high voltage electrodes from the vacuum chamber. The purpose of the goniometer is to control the pitch and yaw of the ion source while under vacuum in order to correct for off-axis emission, as further detailed in Ulibarri, \textit{et al.} \cite{ulibarri2024goniometer}  A polytetrafluoroethylene (PTFE) plate houses a stainless steel 0.125" outer diameter cylindrical reservoir with a porous PTFE press-fit insert for propellant retention, with propellant loading via pipette requiring $\sim$ 5 $\mu$l of RTIL propellant, EMI-BF$_4$. A set screw from the PTFE plate outer edge allows for electrical contact to the reservoir and thus propellant. The stainless steel extractor electrode is affixed 0.15 inches from the reservoir plate, with the needle tip centered within a 0.125 inch diameter hole in the extractor. By applying a high potential difference ($V_{source}$ = 1-3 kV) between the source and extractor under vacuum, a Taylor cone is formed on the tip of the needle. Ion emission begins at a given startup voltage, forming a polydisperse molecular ion plume \cite{petro_multiscale_2022}. In the negative emission mode, a high negative bias is supplied to the ion source via Keithely SourceMeter (2657A) and the extractor is grounded. This allows for the emission of negative ions, mainly in the form of BF$_4^-$ monomers and [BF$_4^-$][EMI-BF$_4$] dimers, with a fraction of the plume existing as [BF$_4^-$][EMI-BF$_4$]$_2$ trimers and higher $m/z$ droplets. Conversely, in the positive emission mode a high positive bias is applied to the source and the plume consists of EMI$^+$ monomers, droplets, and cationated neutrals ([EMI$^+$][EMI-BF$_4$]$_n$ with $n$ = 1 dimers and $n$ = 2 trimers). Emission current magnitudes generally range from 100 nA - 400 nA for the tungsten ion source employed for the following experiments. The plume divergence half angle has been simulated by Petro, \textit{et al.} \cite{petro_multiscale_2022} within 7-12$^{\circ}$ from plume centerline, with experiments showing closer to 20 degrees \cite{Lozano_2006}. The purpose of this ion source design is to allow for adjustment of the firing angle of the tungsten source while under vacuum in order to adjust for off-axis emission by control of the goniometer via flexible shafts. An additional benefit of this design is semi-consistent needle to extractor distance from test-to-test and the mitigation of propellant bubbling during vacuum pump down. The use of a single emitter as compared to thruster arrays allows for more careful correlation of incoming beam properties (current, energy spectra, mass spectra) with secondary species.

\subsection{Target}

The electrospray plume is directed at a target biased to high voltage, $V_{target}$, where impact of the primary plume and a target surface results in a sputtering, or the removal of atoms, molecules, or ions from a surface \cite{van_der_heide_secondary_2014}. The working distance measured approximately 2 inches from the tungsten tip to the surface. For preliminary tests, the target consisted of a 100 mm diameter silicon wafer deposited with a 100 nm of gold on a 10 nm titanium binding layer. These wafers were created via e-beam evaporation with a CVC4500 evaporator at the Cornell NanoScale Facility. Atomically flat, uniform metallic surfaces were chosen for preliminary electrospray TOF-SIMS targets to mitigate inconsistent signals, as SIMS signals are highly sensitive to surface topology and variations in smoothness and uniformity can lead to inconsistent signals and reduce the reliability of the analysis \cite{van_der_heide_secondary_2014}. Gold was selected as the target of interest in the following studies due to its use within thruster systems like extractor grids \cite{petro_characterization_2020}.

The secondary ion acceleration grid consists of a high-transparency stainless steel mesh grid affixed via alumina spacers 0.25 inches from the target surface. This grid is at electrical ground, creating an electric field between the target and grid that both allows for (1) acceleration of the primary ions to the target and, (2) simultaneously extracting secondary ions from the sputtered cloud. An important note about this experimental setup is the polarity of the voltages on the source and target. The target is maintained at a high potential, $V_{target}$, with polarity opposite to the source voltage, $V_{source}$, resulting in a total impact energy equal to the sum of these two voltages. It is important to note that the target bias is modular, allowing $V_{target}$ to be adjusted based on the specific analysis. When $|V_{target}| < |V_{source}|$, like-polarity measurements (positive primary, positive secondary, and vice versa) are enabled. However, this configuration not only reduces impact energies but also the secondary ion accelerating potentials, which in turn increases secondary beam spreading thus reducing secondary yields. All of this combined lowers the intensity and signal-to-noise ratio of TOF measurements. 
\subsection{Time-of-Flight Mass Spectrometer} The time-of-flight mass spectrometer (TOF-MS) system allows for the identification of secondary ion's mass-to-charge ratio. Time-of-flight is a standard electrospray diagnostic for determining the composition of the ion plume for indirect thrust and specific impulse measurements \cite{lozano_ionic_2005, lyne_simple_2023}. The system used for this ESI TOF-SIMS diagnostic replicates a typical in-house system used for electrospray TOF-MS, with the specific system employed in this diagnostic detailed in Ref. \cite{cogan_2023_electrospray}. For electrospray TOF-SIMS, secondary ions are accelerated from the target surface to form an approximately monoenergetic secondary ion plume with energy essentially equivalent to $V_{target}$ given their low ejection energies from the sample. This plume is passed through an electrostatic gate consisting of two parallel electrodes connected to a high voltage power supply and pulsed from 0 V to 950 V at a frequency of 1 kHz via a Keysight EDU33212A waveform generator. The gate acts to alternately deflect and let the plume pass through at this frequency. The secondary plume travels down the TOF tube, with species separating according to mass-to-charge ($m/z$) ratio. Detection occurs at the microchannel plate (Hamamatsu F12107, gain $\sim$ 10$^4$) coupled with a transimpedance amplifier (Advanced Research Instruments Corporation TDC-30, 0.5 V/1 µA) with the signal recorded by an oscilloscope. The flight time, $t_{TOF}$,  is the time it takes for a particle to travel from the center of the gate to the microchannel plate detector. This flight time is related to the mass of the particle, $m$, by conservation of energy, 

\begin{equation}
\label{eq: tofeqn}
    t_{TOF} = \frac{L_{TOF}}{\sqrt{2(\frac{z}{m})V_{target}}} + t_{delay}
\end{equation}

\noindent where $L_{TOF}$ is the length of the flight tube, $z$ is the charge state of the particle, and $t_{delay}$ is the time delay between the oscilloscope trigger and the actual time that the electrostatic gates come to high potential (which is typically on the order of hundreds of nanoseconds). For $\sim$ 1 amu mass resolution, $L_{TOF}$ is approximately 1 m but varied slightly between experiments. The TOF raw signal records current as a function of time, with each step in current indicating a species arriving at the detector \cite{jia-richards_quantification_2022, petro_transient_2019}. These raw signals are averaged $>$ 16,000 times over steady state operation by the oscilloscope. By taking the derivative of the current trace, a spectral representation of the plume composition can be obtained.

\subsubsection{X-Axis Determination}
\label{sec: methods-x axis determination}
Due to the nature of the TOF-SIMS target in the vacuum chamber, the precise flight distance is difficult to accurately measure and changes slightly from experiment to experiment. As this is a critical parameter in Eqn. \ref{eq: tofeqn}, this complicates precise determination of the x-axis of TOF curves and mass spectra. In addition, the electrostatic gate is pulsed to high potential by a square wave from a function generator, which also triggers the data capture via an oscilloscope. There is a time delay, $t_{delay}$, in Eqn. \ref{eq: tofeqn}, of hundreds of nanoseconds between the triggering square wave and the actual time when the gate comes to full potential and the gate opens or closes. This further complicates the x-axis determination.

To enable more precise x-axis determination, Exponentially modified Gaussians (EMGs) were fitted to the smoothed derivative mass lines in each spectrum, similar to the approach detailed further in Ref. \cite{ulibarri_2023_detection} These EMGs have the form 
\begin{equation}
    a_0 \frac{\lambda}{2}e^{\frac{\lambda}{2} \left(2 \mu + \lambda \sigma^2 - 2t \right)} \cdot erfc\left(\frac{\mu +\lambda \sigma^2 -t}{\sqrt{2}\sigma}\right)
\end{equation} 
where $a_0$ is the amplitude, $\lambda$ is an exponential decay term, $\sigma ^2$ is the variance, $\mu$ is the mean, and $erfc$ is the complementary error function. Initial guess EMGs of arbitrary small amplitude and variance were visually placed at the time indices of observed mass lines, and the SciPy curve fitting routine was then used to determine the fitted parameters for each mass line. The index of the local maximum of each resultant EMG was taken as the EMG's location, and two known or suspected mass lines were then assigned mass numbers. A simple minimization was then performed to determine which flight distance and time delay resulted in the smallest mass error for these identified lines. The specific mass lines chosen to constrain the x-axis are discussed in detail in Section \ref{sec: results, SIMS}.

\section{Results}
\label{sec: Results}

\subsection{Ion Source Plume Composition}
\begin{figure}
\includegraphics[width=1\linewidth]{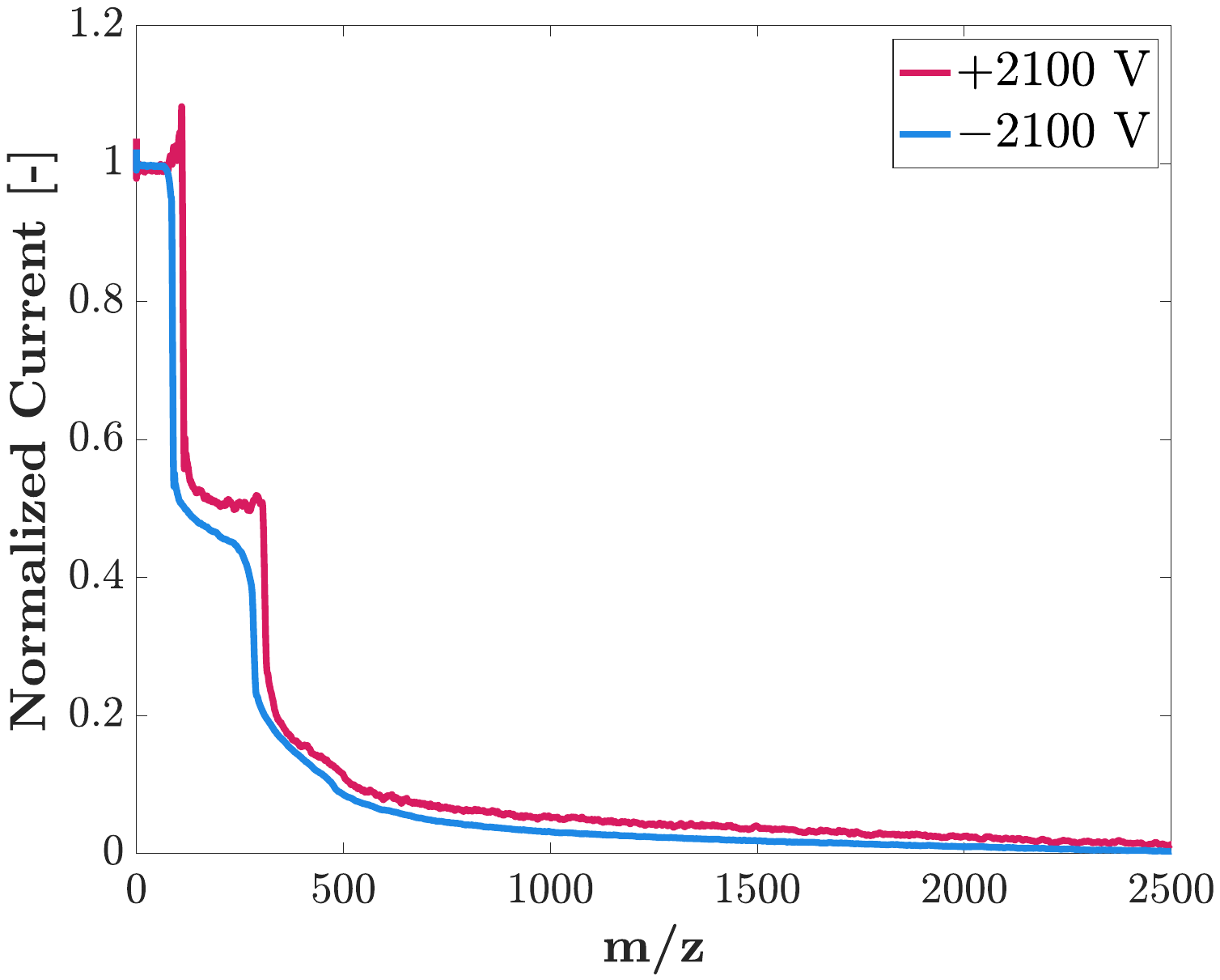}
\caption{\label{fig:primarytof} Primary time-of-flight of both firing modes of the single externally-wetted tungsten ion source. }
\end{figure}

Direct TOF-MS analysis of the primary electrospray plume was performed to evaluate plume composition of the single emitter ion source as shown in Fig.\@ \ref{fig:primarytof}. The raw TOF curves were smoothed with a quadratic Savitzky-Golay filter smoothed over successive 100 data point bins. \cite{savitzky1964smoothing} Firing in the source positive mode with $V_{source}$ at + 2.1 kV and $L_{TOF}$ = 1.01 m resulted in a beam primarily composed of the monomer and dimers of trimers and droplets. These higher $m/z$ droplets to at least $\sim m/z =$ 10$^3$ are shown by the tail of the plot in Fig.\@ \ref{fig:primarytof}. There likely exists even higher $m/z$ species, but the timescales of the data collection prevented visualizing the TOF curve gradient converging to zero \cite{krejci_emission_2017}. This result was paralleled in the negative firing mode with $V_{source}$ at $-$2.1 kV and $L_{TOF}$ = 1.01 m, as shown in Fig.\@ \ref{fig:primarytof}. Both of these confirm that the ion source emits primarily a plume of ions with a subset of higher mass-to-charge ratio droplets mirroring a typical compositions of ion-mode ionic liquid electrospray plumes \cite{jia-richards_quantification_2022, lozano_ionic_2005, krejci_emission_2017}.

\subsection{Secondary Ion Spectra}
\label{sec: results, SIMS}
To demonstrate diagnostic capability, electrospray TOF-SIMS analysis was preformed in both secondary ion polarities with a primary plume impact energy at 4 keV in a \num{5e-5} Torr vacuum environment. For the positive secondary ion analysis, the source voltage, V$_{source}$, was held at $-$2 kV while the sample voltage, $V_{sample}$ was held at +2 kV. Similarly, V$_{source}$, was held at $+$2 kV while the sample voltage, $V_{sample}$ was held at $-$2 kV for the negative secondary ion analysis. For both polarities, the magnitude of the source firing current remained at 250 nA $+/$ 20 nA throughout tests.

X-axis error minimization as detailed in Sec. \ref{sec: methods-x axis determination}, was done with the EMI$^+$ ion at mass 111 and a faint suspected carbon line at mass 12, yielding a flight distance of 0.97 meters and a $t_{delay}$ of 299 ns, both of which are close the expected values based on direct measurement and other unrelated experiments \cite{cogan_2023_electrospray}. Because the parameters should be constant across a single experiment, these positive ion parameters were then copied over to the negative ion spectra, where the lines were fewer and thus harder to identify. In other experiments, the offset has been observed to fluctuate over a range of 100 ns, although it is typically lower than this. Applying such a range to our datasets would result in an error of approximately $\pm$1.4 amu for our highest detected mass at 111 amu. Ongoing work is attempting to understand this fluctuation better so as to reduce this potential source of error.

\subsubsection{Positive Secondary Ion Spectrum}
\label{sec: results, + SIMS}

\begin{figure}[b]
\includegraphics[width=1\linewidth]
{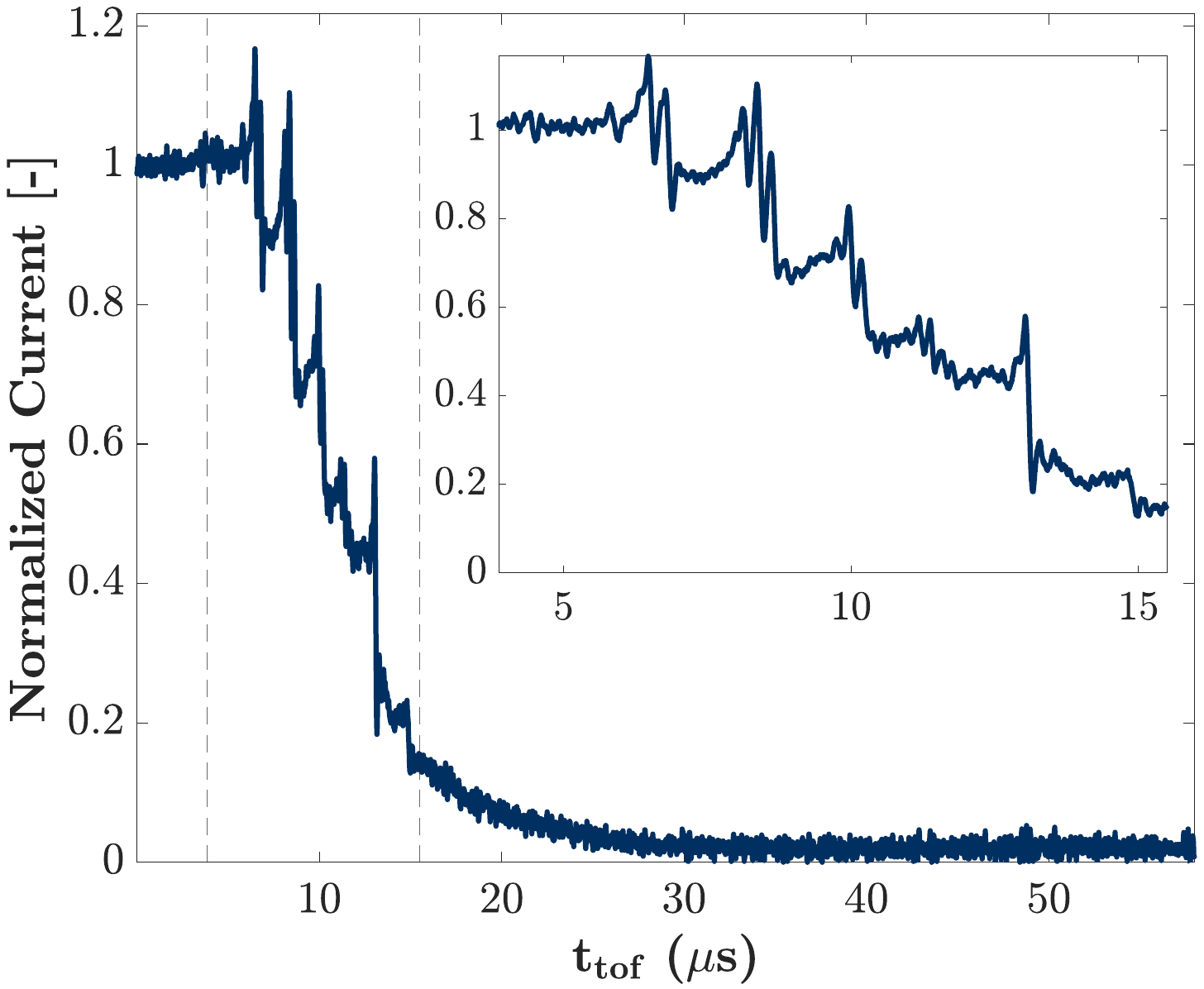}
\caption{\label{fig: raw_au+} Raw time-of-flight measurement of secondary ions produced by 4 keV negative plume impacts with Au target.}
\end{figure}

The raw time-of-flight curve for the positive secondary ions formed from a negative EMI-BF$_4$ plume impacting a gold target is shown in Fig. \ref{fig: raw_au+}. While the figure reports normalized signal, the maximum voltage reading on the oscilloscope read tens of millivolts, approximately one to two orders of magnitude less than typical single volt readings for primary TOF measurements on this TOF system. Given this voltage corresponds to a current signal amplified by $\sim$ 5 x 10$^9$, these voltage readings corresponds to nano to picoamp levels of current for secondary ions reaching the multichannel plate detector. Clearly, there exists ringing within the system corresponding to the spikes before subsequent current drops in the raw data. This is presumably attributed to either impedance mismatching within the high-voltage circuitry or an unstable high voltage connection to the pulsed electrostatic gate. Regardless, the time-of-flight curves show distinct arrival times of species before 20 $\mu$s, with a relatively small droplet tail extending to $\sim$ 30 $\mu$s  before the curve gradient converging to zero. An expanded view of the earlier species detection to 15 $\mu$s is shown in the inset, which shows six distinct groupings of species detection. Given the ringing, there appears to be a spike before a species detection. Therefore, the first four groups seems to have multiple species detection within a relatively short time span, corresponding to multiple species within a small mass range.

\begin{figure*}
\includegraphics[width=.75\linewidth]{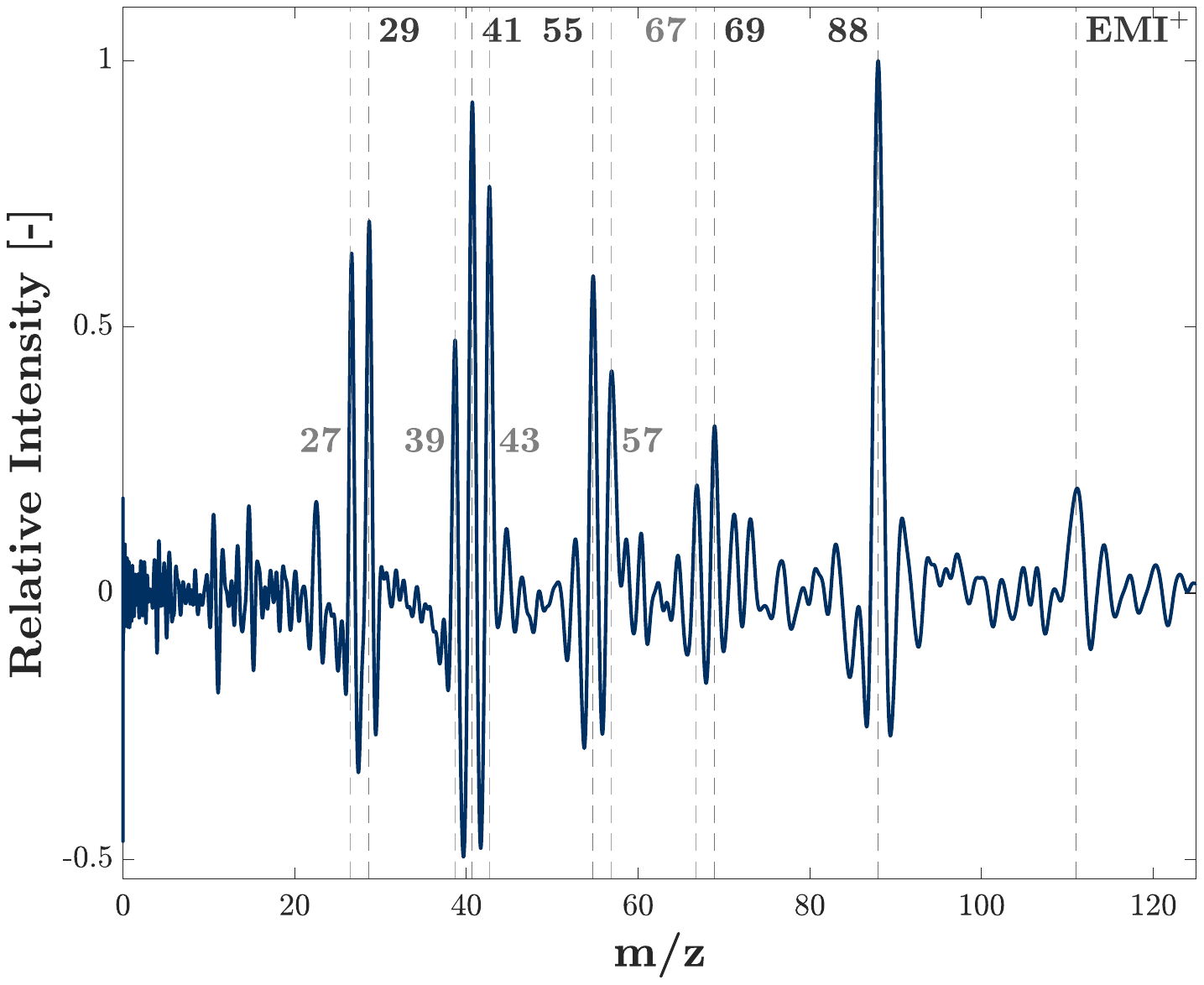}
\caption{\label{fig: au+} Positive secondary mass spectra from a negative EMI-BF$_4$ plume impacting with Au target. Relative most intense peaks are bolded.}
\end{figure*}

Figure \ref{fig: au+} represents the derivative of the raw time-of-flight curve of the positive secondary ions with noise smoothing as detailed in Section \ref{sec: methods-x axis determination}. Preliminary identification of the detected peaks are found in Table \ref{tab:positive species}, with references that refer to works that have also detected the given peak in a RTIL related study. All reported potential peaks have been validated as chemically viable in the NIST Chemical Webbook \cite{nistwebbook}.

A notable characteristic is the presence of multiple ``families' of detected species, which are characteristic of organic-related fragmentation patterns. Each familial peak corresponds to a methyl group loss from the most massive group ($m/z$ = 69 amu) to the least ($m/z$ = 27 amu), with peaks within a few amu within each family. Chemically, this represents molecules with similar structural compositions. The most massive peak at $m/z$ = 111 amu most likely corresponds to the EMI$^+$ cation, which is found as the most dominant positive ion peak in commercial TOF-SIMS of EMI-containing RTIL thin films \cite{bundaleski_ion-induced_2013, GUNSTER20083403}. The relative lower intensity of this peak potentially indicates deposition of propellant on the target surface that is re-sputtered as seen by \citet{fujiwara_time--flight_2014}, given that these high impact-energy collisions are not as likely to produce intact EMI cations \cite{bendimerad_molecular_2022}. The most intense peak, $m/z$ = 88 amu does not possess a surrounding spectral family, and is > 5 amu from typical ionic liquid fragments 83 amu and 96 amu \cite{bundaleski_ion-induced_2013, GUNSTER20083403, shaik_characterization_2024, stefan_CID_2024}. Possible species include the oxalate cation (C$_2$O$_4^+$) arising from oxide and hydrocarbon layers from the target surface or carbon tetraflouride cation (CF$_4^+$) given the presence of carbons and flourines from the impacting BF$_4^-$ monomer, but remains largely unknown.

\begin{table}
\caption{\label{tab:positive species} Positive secondary ion collection mode species detection, with mass resolution +/$-$ 1.4 amu.}
\begin{ruledtabular}
\begin{tabular}{ccc}  Mass (amu) \footnote{Assuming singly charged species.}& Potential Species & Reference Detection \\ \hline
  27 & C$_2$H$_3^+$  & [ \hspace{-0.5em} \citenum{bundaleski_ion-induced_2013},\citenum{GUNSTER20083403}]\\
  29 & C$_2$H$_5^+$, CH$_3$N$^+$, CHO$^+$ & [ \hspace{-0.5em} \citenum{bundaleski_ion-induced_2013}, \citenum{GUNSTER20083403}, \citenum{stefan_CID_2024}]\\
  ~  & ~ & ~ \\
  39 & C$_3$H$_3^+$ & ~ \\
  41 & C$_3$H$_5^+$, C$_2$H$_3$N$^+$  & [ \hspace{-0.5em} \citenum{bundaleski_ion-induced_2013}, \citenum{stefan_CID_2024}] \\
  43 & C$_3$H$_7^+$,  C$_2$H$_3$O$^+$ & [ \hspace{-0.5em} \citenum{bundaleski_ion-induced_2013}] \\
  ~  & ~ & ~ \\
  55 & C$_4$H$_7^+$, C$_3$H$_6$N$^+$, CN$_2$O$^+$ & [ \hspace{-0.5em} \citenum{bundaleski_ion-induced_2013}, \citenum{stefan_CID_2024}] \\
  57 & C$_4$H$_9^+$, C$_3$H$_5$O$^+$ & ~ \\
  ~  & ~ & ~ \\
  67 & C$_4$H$_6$N$^+$ & [ \hspace{-0.5em} \citenum{stefan_CID_2024}]\\
  69 & C$_5$H$_9^+$, C$_3$N$_2$H$_5^+$, CF$_3^+$, C$_4$H$_5$O$^+$ & [ \hspace{-0.5em} \citenum{bundaleski_ion-induced_2013}, \citenum{GUNSTER20083403}] \\
   ~ & ~ & ~ \\
  88 & C$_2$O$_4^+$, CF$_4^+$ & ~ \\
111  & EMI$^+$ & [ \hspace{-0.5em} \citenum{bundaleski_ion-induced_2013}, \citenum{GUNSTER20083403}, \citenum{stefan_CID_2024}] \\
\end{tabular}
\end{ruledtabular}
\end{table}

The four spectral families with relative most intense peaks $m/z$ = 29, 41, 55, and 69 amu are typically seen in many organic mass spectra \cite{arisz2020dynamics}, including commercial TOF-SIMS of ionic liquids with imidazole cations \cite{bundaleski_ion-induced_2013, GUNSTER20083403}, residual gas analyzer studies of neutral particles produced by electrospray plume impacts with stainless steel \cite{shaik_characterization_2024}, and collision-induced disassociation studies of EMI-BF$_4$. \cite{stefan_CID_2024} \citet{bundaleski_ion-induced_2013} in commercial TOF-SIMS of an EMI-Im film report masses $m/z$ = 27, 29 as the ethyl alkyl functional group fragments (C$_2$H$_3^+$, C$_2$H$_5^+$), as well as $m/z$ = 41, 55 as two bond scissions of the imidazolium ring (C$_2$H$_3$N$^+$, C$_3$H$_6$N$^+$ respectively). Protonated imidazolium rings correspond to $m/z$ = 69 amu (C$_3$H$_5$N$_2^+$) with the imidazolium ring cation itself equivalent to $m/z$ = 68 amu (C$_3$H$_4$N$_2^+$) and within the mass resolution of the data to be represented by the $m/z$ = 67 or 69 peak \cite{bundaleski_ion-induced_2013, GUNSTER20083403}. Therefore, almost all peaks have been observed in fragmentation studies of EMI-containing RTILs. 

These peaks also appear as fragments of hydrocarbon containing materials. It is known that surfaces without proper decontamination and mitigation techniques posses layers of hydrocarbons and oxides \citet{klosterman_ion-induced_2022}. describe nanometer thick oxide and hydrocarbon layers on gold surfaces. This, combined with the relatively high vacuum pressure contributing to adsorbate layer formation due to the Hertz-Knudsen relationship \cite{van_der_heide_secondary_2014} and pump oil being a known contaminant of TOF-SIMS samples \cite{stipdonk_1999_NaBF4} gives a likely probability that these peaks are as a result from hydrocarbon secondary ions unrelated to the ionic liquid plume. For instance, peaks $m/z$ = 41, 43, 55, 57, 69 all could be hydrocarbon-related secondary ions  (C$_n$H$_m^+$, n > 2) that could not be formed from the fragments of an EMI$^+$ cation's ethyl functional group. These spectral patterns have been observed in TOF-SIMS of crude oil \cite{arisz2020dynamics, siljestrom2009}, as well as in studies observing backstreaming vacuum pump oil \cite{tsutsumi_prevention_1990, maurice_oil_1979}. In summary, the detected secondary peaks align both with typical fragments of the primary molecule, EMI$^+$, and with typical cracking patterns of organic hydrocarbons that may exist on outermost layer of target's surface.

\subsubsection{Negative Secondary Ion Spectrum}
\begin{figure}
\includegraphics[width=1\linewidth]{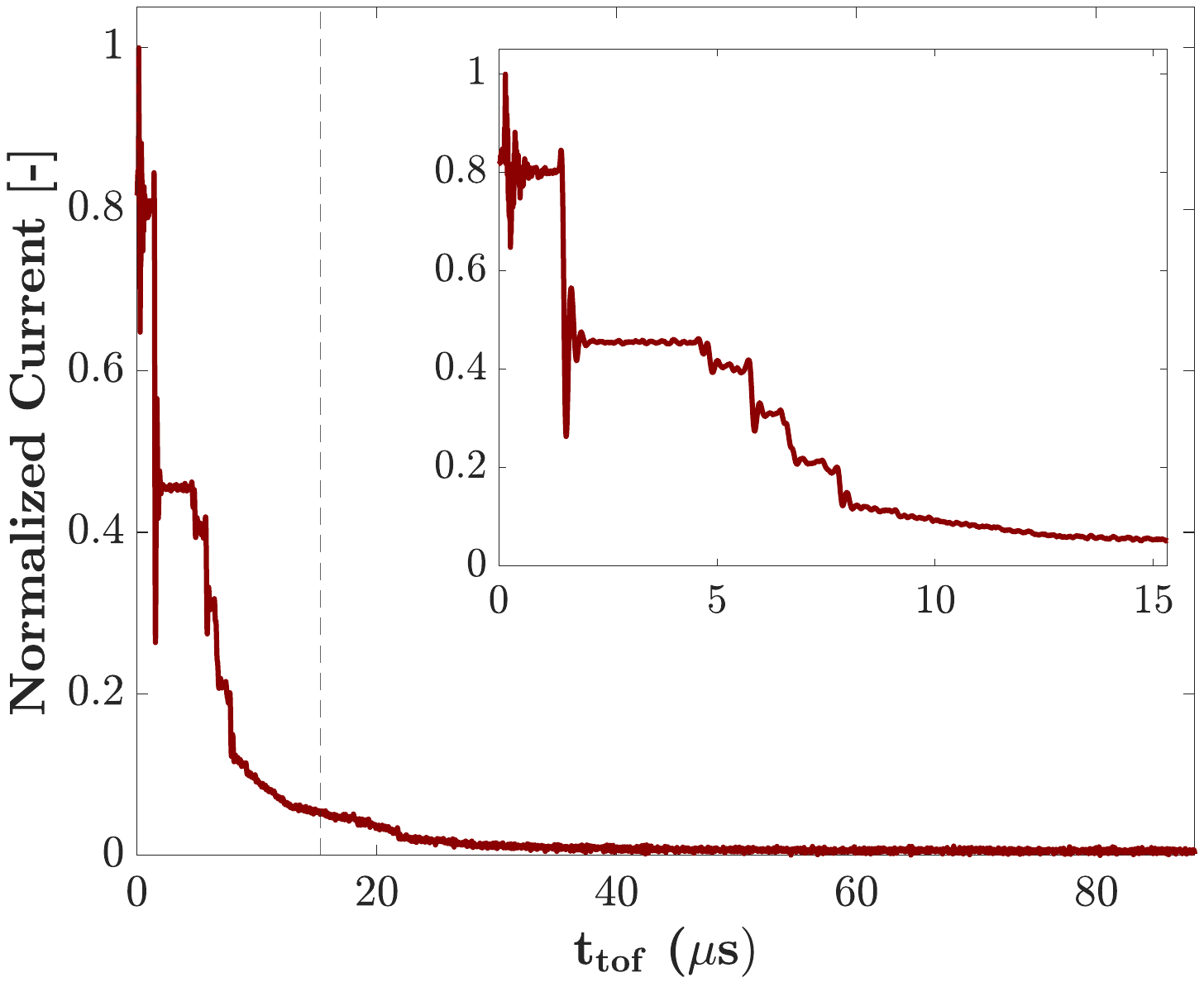}
\caption{\label{fig: raw_au-} Raw time-of-flight measurement of negative secondary ions produced by 4 keV positive plume impacts with Au target.}
\end{figure}

The raw time-of-flight curve for the negative secondary ions formed from a positive EMI-BF$_4$ plume impacting a gold target is shown in Fig. \ref{fig: raw_au-}. The raw oscilloscope voltage readings corresponding to amplified current signals showed a peak of several hundred millivolts for the negative secondary ions, with an average signal strength 2.8 times higher than that of the positive secondary ions (0.15 V vs. 0.053 V). Again, the electrical ringing is present but again exhibits the same pattern of signal spikes at a given species arrival time. Compared to the positive raw spectra in Fig. \ref{fig: raw_au+}, the negative secondary ions are lighter, with the bulk molecular ions having flight times less than 10 $\mu$s. Negative heavier droplets extend to a similar mass range as the positive spectra, with the population extending to $\sim$ 30 $\mu$s.  An expanded view of the lighter species detection to 15 $\mu$s is shown in the inset, which again shows six distinct groupings of species detection. An important note is the decreased mass resolution for the negative spectrum compared to positive due to inconsistent oscilloscope data window capture. 

\begin{figure*}
\includegraphics[width=.75\linewidth]{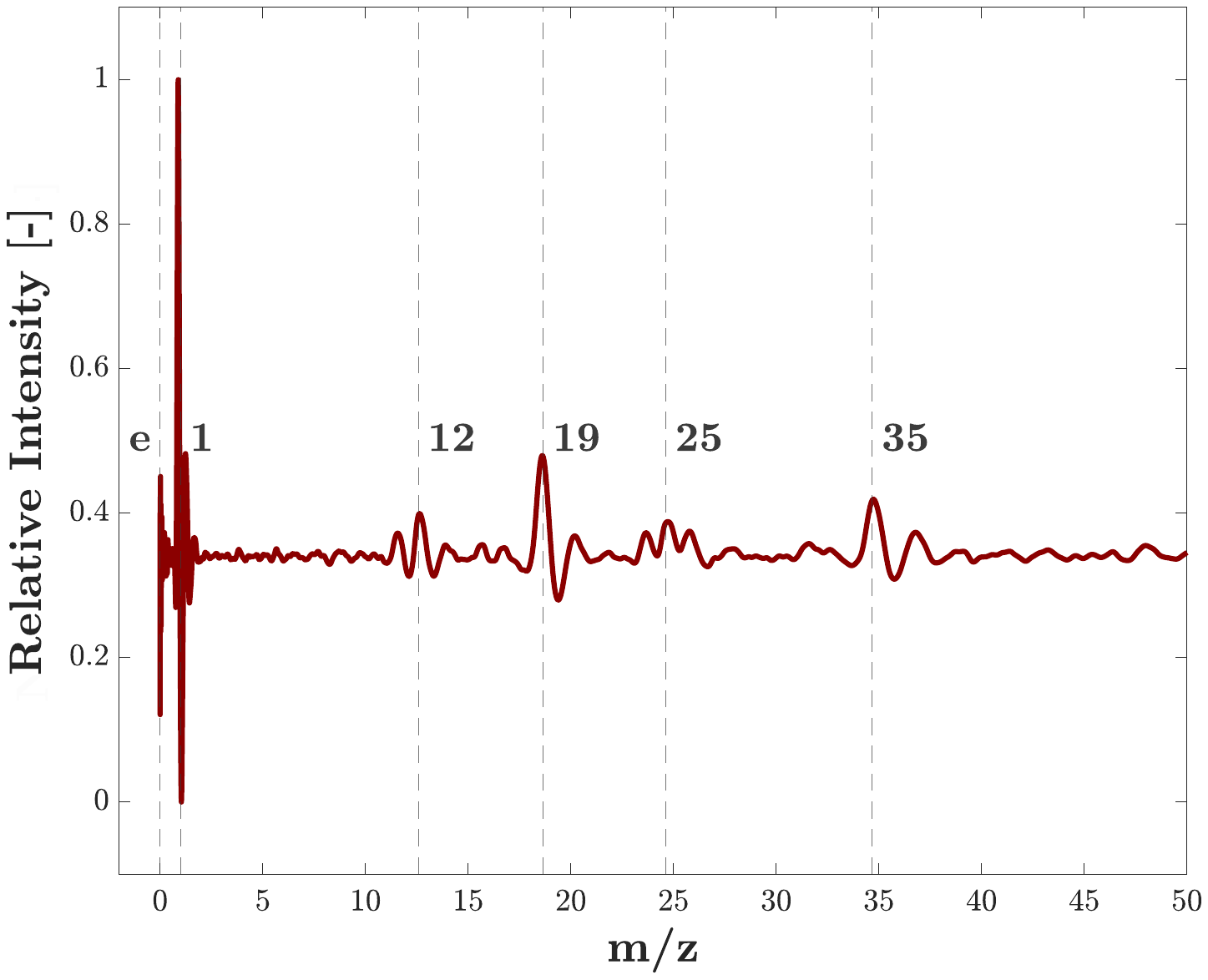}
\caption{\label{fig: au-} Negative secondary mass spectra from a positive EMI-BF$_4$ plume impacting a gold substrate.}
\end{figure*}

The derivative of the smoothed time-of-flight curve to give a typical spectral representation of the negative secondary ions is shown in Fig. \ref{fig: au-}. The sprectrum has an intense peak at $m/z$ = 1 amu, or the hydrogen anion (H$^-$, H$_2^-$) as seen in other organics and RTIL related works. Followed are much less intense peaks at $\sim$ $m/z$ = 12, 19, 25 and 35 amu. \citet{bundaleski_ion-induced_2013} reports a negative ions from EMI-Im film from the spectral family around $m/z$ = 25 amu, with C$_2^-$ ($m/z$ = 24),  C$_2$H$^-$ ($m/z$ = 25), and C$_2$H$_2^-$ + CN$^-$ ($m/z$ = 26) which is reflected in this spectra, reportedly as fragmentation products from the primary EMI$^+$ cation. These carbon-containing species were also reported in \citet{stipdonk_1999_NaBF4} in the TOF-SIMS study of sodium tetrafluoroborate (NaBF$_4$), but as products of oil backstreaming contamination. The $m/z$ = 19 amu peak most closely aligns with the flourine anion, as seen in both TOF-SIMS of EMI-Im \cite{bundaleski_ion-induced_2013, GUNSTER20083403} and numerical reactive force field molecular dynamics simulations of EMI-BF4 impacting potential walls \cite{bendimerad_molecular_2022, shaik_characterization_2024}. Given the mass resolution of +/$-$ 1.4 amu, products like hydroflouric acid (HF$^-$) may also be represented in the $m/z$ = 19 peak. The peak observed at $m/z = 35$ does not correspond to any immediately identifiable species. However, similar peaks within this mass range have been reported in previous studies, such as those by \citet{bundaleski_ion-induced_2013} and \citet{stipdonk_1999_NaBF4}, and possible species assignment is found in Table \ref{tab:negative species}.

\begin{table}
\caption{\label{tab:negative species} Negative Secondary Ion collection mode species detection, with mass resolution +/$-$ 1.4 amu.}
\begin{ruledtabular}
\begin{tabular}{ccc}  Mass (amu) \footnote{Assuming singly charged species.}& Potential Species & Reference Detection \\ \hline
  0 & Electrons (e$^-$) & ~\\
  1 & H$^-$, H$_2^-$ & [ \hspace{-0.5em} \citenum{bundaleski_ion-induced_2013}]\\
  ~  & ~ & ~ \\
  12 & C$^-$, CH$^-$ & [ \hspace{-0.5em} \citenum{stipdonk_1999_NaBF4}] \\
  19 & F$^-$, HF$^-$ & [ \hspace{-0.5em} \citenum{bundaleski_ion-induced_2013}, \citenum{GUNSTER20083403}] \\
  ~  & ~ & ~ \\
  25 & C$_2^-$, C$_2$H$^-$, C$_2$H$_2^-$, CN$^-$ & [ \hspace{-0.5em} \citenum{bundaleski_ion-induced_2013}, \citenum{stipdonk_1999_NaBF4}] \\
  ~  & ~ & ~ \\
  35 & H$_2$O$_2^-$, B$_2$N$^-$ & ~ \\
\end{tabular}
\end{ruledtabular}
\end{table}


\section{Conclusion}
\label{sec: conclusion}

A novel electrospray time-of-flight secondary ion mass spectrometry diagnostic was experimentally validated to probe secondary ion mass-to-charge ratio chemical composition from plume-surface impacts. This was achieved via a single tungsten electrospray ion source operating with room temperature ionic liquid propellant, EMI-BF$_4$, in a $\sim$ 10$^{-5}$ Torr vacuum. Impacts of the primary plume with gold metallic targets induced secondary ion emission characterized by time of flight mass spectrometry. Positive secondary ion mass spectra generated from primary negative plume impacts showed potential ionic and covalent dissociation of the primary beam given the preliminary detection of EMI$^+$ and related fragmentation species. In addition, the positive spectrum indicated multiple species of $m/z <$ 100 amu, pointing to either potential primary ion impact fragmentation products or to adsorbed hydrocarbon layers -- or a combination of both. In the negative secondary ion spectrum, the signal was dominated by H$^-$ ions. However, multiple ion species of $m/z <$ 40 amu were detected, again most likely related to hydrocarbon and primary ion impact fragmentation products. For both spectra, the detected species are of significant interest as they are the result of the opposite polarity primary plume impacts. These species are the most likely to backstream and contribute to lifetime-limiting processes.

It is important to note that this diagnostic can be replicated using standard instruments commonly found in electric propulsion laboratories, allowing for the study of a wide range of plume-surface interactions. While the described diagnostic utilizes a microchannel plate detector, a current-collecting electrode paired with a transimpedance amplifier that offers sufficient gain and a strong signal-to-noise ratio could also be effective, as demonstrated in systems like \citet{lyne_simple_2023}. Additionally, this diagnostic is source-agnostic, meaning the single emitter in this study can be replaced with emitter arrays and thrusters, or different types of emitters across the spectrum of electrospray variants (such as colloidal and porous emitters). Since surface processes vary based on the primary plume composition \cite{uchizono_role_2021}, the resulting secondary species composition is likely to differ depending on the source and target composition. This flexibility is particularly crucial for facility calibrations, as it enhances the understanding of the secondary species populations generated from plume interactions with vacuum chamber surfaces and how those secondaries ions influence electrospray operation as well as other diagnostics. 

Future work will focus on determining the source of secondary ions -- whether from target-based contamination or from primary plume fragmentation. This information will be critical in determining relevant secondary ion species in terms of intrinsic thruster operation or facility effects. In addition, it will be important to also characterize like-polarity secondary ions as the primary plume to understand secondary ion generation. This work can then be applied to test the permutation of thruster operating conditions, propellants, and surfaces of interest for a comprehensive electrospray TOF-SIMS analysis of electrospray thruster plume-surface interactions.

Electrospray propulsion devices show promise to be mission-enabling technologies. Characterizing secondary ions that form via plume-surface interactions will enable both high-fidelity flight qualification tests and thruster operations, as well as inform secondary species mitigation techniques.

\section{Acknowledgments}
This work was supported by a NASA Space Technology Graduate Research Opportunity fellowship (80NSSC23K1212). This work was supported in part by the Heising Simons Foundation 51 Pegasi B Fellowship grant number 2024-5175. 
A portion of this work was supported by a grant from the Jet Propulsion Laboratory, California Institute of Technology, under contract with the National Aeronautics and Space Administration (80NM0018D0004).

\section{Data Availability}
The data that support the findings of this study are available from the corresponding author upon reasonable request.

\appendix

\nocite{*}
\bibliography{aipsamp}

\end{document}